\documentclass[preprint2]{aastex6}
\usepackage{color}
\usepackage[titletoc]{appendix}
\usepackage[fleqn]{amsmath}
\usepackage{mathtools}
\usepackage{float}
\usepackage{comment}
\usepackage{enumitem}
\usepackage{natbib}
\usepackage{bm}

\def\approxprop{%
  \def\p{%
    \setbox0=\vbox{\hbox{$\propto$}}%
    \ht0=0.6ex \box0 }%
  \def\s{%
    \vbox{\hbox{$\sim$}}%
  }%
  \mathrel{\raisebox{0.7ex}{%
      \mbox{$\underset{\s}{\p}$}%
    }}%
}

\shortauthors{Millholland \& Batygin}
\shorttitle{Primordial Obliquities}

\begin{document} 

\title{Excitation of Planetary Obliquities Through Planet-Disk Interactions} 
\author{Sarah Millholland$^{1,3}$ and Konstantin Batygin$^{2}$}
\affil{$^1$ Department of Astronomy, Yale University, New Haven, CT 06511, USA \\ 
$^2$ Division of Geological and Planetary Sciences, California Institute of Technology, Pasadena, CA 91125, USA}
\altaffiliation{$^3$ NSF Graduate Research Fellow}
\email{sarah.millholland@yale.edu}

\begin{abstract}
The tilt of a planet's spin axis off its orbital axis (``obliquity'') is a basic physical characteristic that plays a central role in determining the planet's global circulation and energy redistribution. Moreover, recent studies have also highlighted the importance of obliquities in sculpting not only the physical features of exoplanets but also their \textit{orbital} architectures. It is therefore of key importance to identify and characterize the dominant processes of excitation of non-zero axial tilts. Here we highlight a simple mechanism that operates early on and is likely fundamental for many extrasolar planets and perhaps even Solar System planets. While planets are still forming in the protoplanetary disk, the gravitational potential of the disk induces nodal recession of the orbits. The frequency of this recession decreases as the disk dissipates, and when it crosses the frequency of a planet's spin axis precession, large planetary obliquities may be excited through capture into a secular spin-orbit resonance. We study the conditions for encountering this resonance and calculate the resulting obliquity excitation over a wide range of parameter space. Planets with semi-major axes in the range $0.3 \ \mathrm{AU} \lesssim a \lesssim 2 \ \mathrm{AU}$ are the most readily affected, but large-$a$ planets can also be impacted. We present a case study of Uranus and Neptune and show that this mechanism likely cannot help explain their high obliquities. While it could have played a role if finely tuned and envisioned to operate in isolation, large-scale obliquity excitation was likely inhibited by gravitational planet-planet perturbations.

\end{abstract}

\section{Introduction}

The spin dynamics of planetary bodies influence many of their salient features, including atmospheric circulation, climate, and tidal evolution. 
The two most important features of the spin state are the rotation period and the obliquity, defined as the angle between the planet's spin and orbital axes. The present-day obliquities of the Solar System planets are wide-ranging, from near-zero spin-orbit misalignment for Mercury, to $23^{\circ}$ for Earth, and $98^{\circ}$ for Uranus. Notably, five of the eight planets have obliquities larger than $20^{\circ}$.  

Obliquities of extrasolar planets have not yet been measured directly.\footnote{Note these obliquities are distinct from the \textit{stellar} obliquity, which has been constrained for many systems \citep{2015ARA&A..53..409W}.} There are several prospects, however, for measuring obliquities in coming years. Ground-based, high-resolution spectroscopy permits measurements of the $v\sin i$ of the planet, offering a degenerate constraint on the spin rate and component of obliquity along the line-of-sight \citep{2014Natur.509...63S, 2018NatAs...2..138B}. Alternatively, measurements of planetary asphericity from transit photometry offer potential constraint on the spin rate, obliquity, and the coefficient of the gravitational quadrupole moment \citep{2002ApJ...572..540H, 2002ApJ...574.1004S, 2003ApJ...588..545B, Ragozzine2009,  2010ApJ...709.1219C, 2010ApJ...716..850C, 2014ApJ...796...67Z, 2017AJ....154..164B}. 

For close-in planets, features in full-phase, optical or infrared light curve observations may point toward non-zero obliquities \citep{2016MNRAS.457..926S, 2017ApJ...846...69R, 2019ApJ...874....1O, 2019ApJ...874....2O, AdamsMillhollandLaughlin2019}. Similarly, the detection of seasonal variability from infrared photometry or direct imaging has also been proposed as a means of constraining obliquity \citep{2004NewA...10...67G, 2016ApJ...822..112K, 2017AJ....154..204K}. A large obliquity may also be inferred indirectly by considering its effect on other planetary properties, such as its tidal dissipation rate \citep{2018ApJ...869L..15M, 2019NatAs.tmp..218M}.  

Despite the lack of direct detection thus far, there are theoretical expectations for the frequent occurrence of large  planetary obliquities. Several means exist for exciting obliquities from their primordial values, and the Solar System planets are prime examples. First, collisions with massive proto-planetary cores, such as the Moon-forming giant impact, can produce large axial tilts. Similarly, giant impacts are routinely invoked in the origin scenarios of the extreme Uranian tilt \citep{1966SvA.....9..987S, 2012Icar..219..737M}. It is also possible that the obliquities of the giant planets were excited by an early process that twisted the total angular momentum vector of the Solar System, such as asymmetric infall during the system's initial collapse and formation \citep{1991Icar...89...85T}.

Another important obliquity-exciting mechanism is spin-orbit resonance -- a natural syncing of the frequencies of a planet's spin axis precession and orbital precession. Orbital precession is inevitably driven by planet-planet interactions, and spin-orbit resonances of this type likely explain Jupiter's $3^{\circ}$ and Saturn's $27^{\circ}$ obliquities through interactions with Uranus and Neptune, respectively \citep{Ward2004, 2004AJ....128.2510H, 2006ApJ...640L..91W, 2015ApJ...806..143V, 2015AJ....150..157B, 2018ARA&A..56..137N}. Furthermore, in the case of extrasolar planets, \cite{2019NatAs.tmp..218M} showed that enhanced tidal dissipation generated by obliquity excitations in spin-orbit resonances may be a key factor in sculpting the orbital period distribution of systems with multiple transiting planets. 

We note further that when several spin-orbit resonances overlap, large-amplitude, chaotic excursions of the obliquity can result. This has likely been important for all of the Solar System terrestrial planets \citep{1993Natur.361..608L}, especially Venus \citep{2003Icar..163....1C, 2003Icar..163...24C} and Mars \citep{1973Sci...181..260W, 1993Sci...259.1294T}, where the obliquity varies chaotically between $\sim10^{\circ}$ and $50^{\circ}$ on timescales of hundreds to thousands of years. 

Although planet-planet interactions are often considered as a driver of orbital precession, they are not unique. In fact, any deviation from a purely $\propto 1/r$ potential will give rise to nodal recession. Natural sources that act as the system is still forming include the gravitational field of a protoplanetary disk and the quadrupole potential of a rapidly-rotating young star. The precession frequencies associated with these torques decrease in magnitude as the nebular gas dissipates and the young star contracts. If these evolutions result in adiabatic spin-orbit resonance crossings where the spin and orbital precession frequencies are equal, then obliquities will inevitably be excited. 

\begin{figure}
\epsscale{1.3}
\plotone{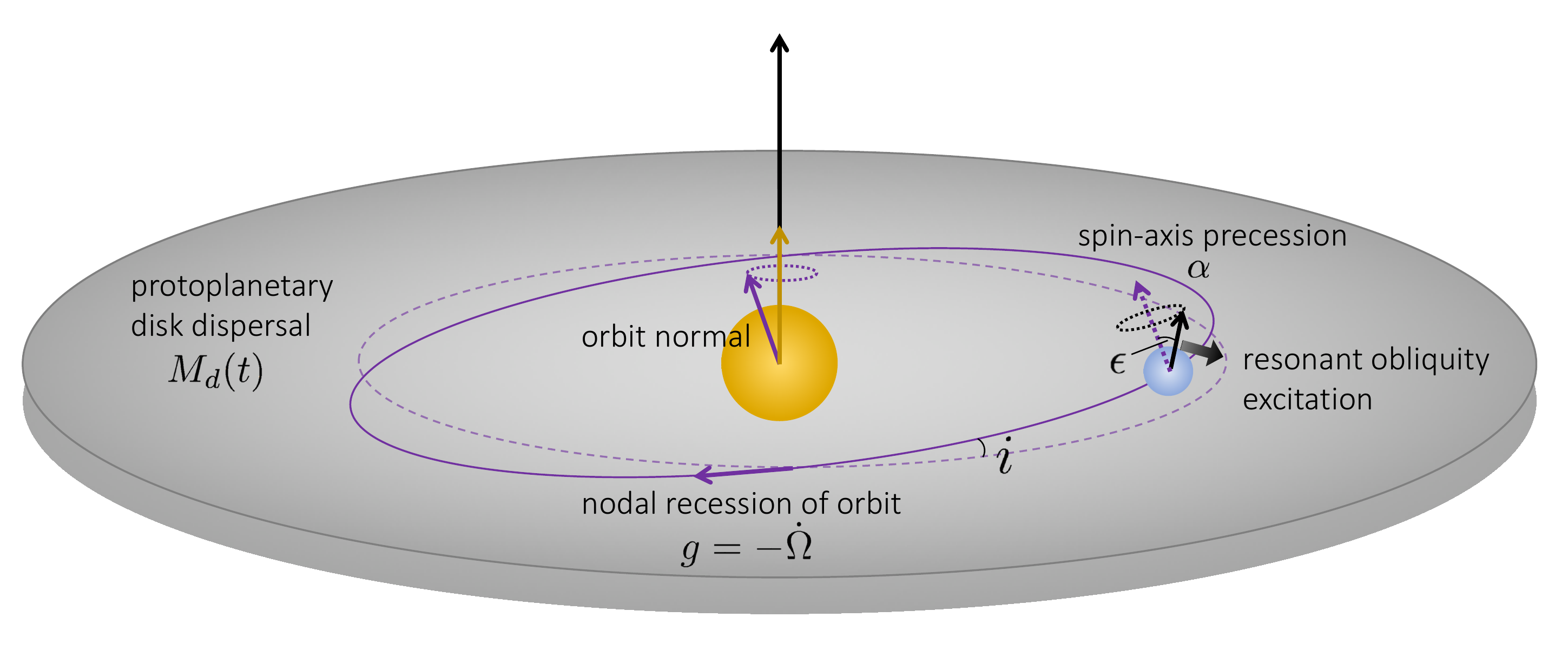}
\caption{A schematic representation of obliquity excitation via capture into disk-induced secular spin-orbit resonance. The protoplanetary disk generates nodal recession of the planet's orbit at frequency $g$. As the disk mass, $M_d(t)$, decreases over time, $g$ reaches a commensurability with the planet's spin axis precession rate, which results in slow resonant tilting of the spin axis away from the orbital axis.}
\label{geometric diagram}
\end{figure}

In this work, we focus on the first of these two torque generators: the protoplanetary disk. We present an explication of the role of disk-induced orbital precession in exciting planetary obliquities. A schematic representation of this process is shown in Figure \ref{geometric diagram}. As we show below, the mechanism is extensive and robust due to the large magnitude of the initial disk-induced precession rate and the wide range of commensurabilities that are swept. However, it has not yet been widely appreciated in the literature\footnote{\cite{2005ApJ...628L.159W} discussed disk-induced secular spin-orbit resonance in the context of hot Jupiters. In this same context, \cite{2007ApJ...665..754F} examined the role of stellar quadrupole-induced resonances. \cite{2019NatAs.tmp..218M} also touched on the influence of a young stellar quadrupole in short-period, compact ``\textit{Kepler}-like'' systems.}.  

It is enlightening to point out that this mechanism is closely analogous to other secular resonances that might occur early on in a system's lifetime. One theory for the excitation of \textit{stellar} spin-orbit misalignments in extrasolar systems involves a resonant commensurability between stellar spin axis precession and nodal regression of a circumstellar disk in a star-disk-binary system \citep{2012Natur.491..418B, 2013ApJ...778..169B, 2014MNRAS.440.3532L, 2014ApJ...790...42S, 2018MNRAS.478..835Z}. Specifically, the stellar spin precesses about the disk angular momentum vector. Meanwhile, the disk angular momentum vector precesses due to a binary star perturber. As the disk dissipates, a resonance ensues when the spin precession and disk precession frequencies match, resulting in adiabatic excitation of the stellar obliquity. This scenario is mathematically similar to the one at hand, except we are now considering the precession of the planetary (rather than stellar) spin about the orbital (rather than disk) angular momentum vector, and this vector itself precesses due to the disk (as opposed to the binary perturber). 

This paper is organized as follows. In Section \ref{section 2}, we define the frequencies of spin axis precession and disk-induced orbital precession. We then use these expressions in Section \ref{section 3} to define analytic criteria for spin-orbit resonance crossing and capture. In Section \ref{section 4}, we develop a simple perturbative model to study the evolution of a planet's obliquity when this resonance is encountered, and map the resulting obliquity excitation over a wide range of parameter space. Finally, we present a special case study of the Uranian and Neptunian obliquities in Section \ref{section 5} and conclude in Section \ref{section 6}.

\section{Spin and Orbital Precession}
\label{section 2}

In order to study the evolution of the spin and orbital precession into resonant commensurability, we must first define their characteristic frequencies. 

\subsection{Spin axis precession}
The torque from the host star on a rotationally-flattened planet causes the planet's spin axis to precess about its orbit normal. The period of precession is 
\begin{equation}
T_{\alpha} = 2\pi/(\alpha\cos\epsilon),
\end{equation} 
where $\epsilon$ is the planet's obliquity (angle between the spin and orbital axes) and $\alpha$ is the precessional constant. The magnitude of $\alpha$ may be strongly enhanced by the presence of close-in satellites or a primordial circumplanetary accretion disk. This more rapid precession is due to the adiabatic gravitational coupling between the oblate planet and surrounding disk/satellites, which makes the system precess as a unit \citep{1965AJ.....70....5G}. 

In the absence of a disk or satellites, $\alpha = \alpha_0$ is given by \citep{Ward2004}
\begin{equation}
\alpha_0 = \frac{k_2}{2C}n f_{\omega}\left[\frac{M_{\star}}{M_p}\left(\frac{R_p}{a}\right)^3\right]^{\frac{1}{2}}.
\label{alpha0}
\end{equation}
Here $M_{\star}$ is the host star mass, $M_p$ and $R_p$ the planet mass and radius, $a$ the semi-major axis of the planet, and $n = 2\pi/P$ its mean motion. The quantity $k_2$ is the planet's Love number, a dimensionless value related to the planet's central concentration of the density profile and its deformation response to tidal disturbance. The quantity $C$ is the planet's moment of inertia normalized by $M_p {R_p}^2$. Finally, $f_{\omega} = \omega/\omega_b$ is defined as the planet's spin rate as a fraction of break-up speed, $\omega_b = \sqrt{G M_p/{R_p}^3}$. The purpose of this parameterization will become clear later. The definition of $\alpha_0$ has incorporated the following form of $J_2$, the coefficient of the quadrupole moment of the planet's gravitational field \citep{Ragozzine2009},
\begin{equation}
J_2 = \frac{\omega^2 {R_p}^3}{3 G {M_p}}k_2.
\end{equation}


As already mentioned above, if the planet is encircled by a circumplanetary disk or a massive satellite, the effective value of $\alpha$ will be larger. If we define $f_{\alpha} = \alpha/\alpha_0$ as the enhancement factor of the precessional constant over its value when there is no disk/satellite, then \citep{Ward2004}
\begin{equation}
\alpha = f_{\alpha}\alpha_0 = \frac{3}{2}\frac{n}{f_{\omega}}\left[\frac{(k_2 {f_{\omega}}^2/3) + q}{C + l}\right]\left[\frac{M_{\star}}{M_p}\left(\frac{R_p}{a}\right)^3\right]^{\frac{1}{2}}.
\label{alpha}
\end{equation}
The quantity $q$ is the effective quadrupole coefficient of the satellite system or disk, and $l$ is the angular momentum of the satellites/disk normalized by $M_p{R_p}^2\omega$. In the case of a single equatorial satellite with mass, $m_s$, and semi-major axis, $a_s$, $q$ and $l$ are defined by
\begin{equation}
\begin{aligned}
q &= \frac{1}{2}\left(\frac{m_s}{M_p}\right)\left(\frac{a_s}{R_p}\right)^2 \\
l &= \left(\frac{1}{f_{\omega}}\right)\left(\frac{m_s}{M_p}\right)\left(\frac{a_s}{R_p}\right)^{\frac{1}{2}}.
\label{q and l}
\end{aligned}
\end{equation}
Similarly, for a circumplanetary disk with mass, $m_{cp}$, radius, $r_{cp}$, and surface density profile, $\Sigma_{cp} = \Sigma_{cp,0}(r/R_p)^{-\gamma}$ with $\gamma < 2$, $q$ and $l$ are defined,
\begin{equation}
\begin{aligned}
q &= \frac{1}{2}\left(\frac{2-\gamma}{4-\gamma}\right)\left(\frac{m_{cp}}{M_p}\right)\left(\frac{r_{cp}}{R_p}\right)^2  \\
l &= \left(\frac{2-\gamma}{5/2-\gamma}\right)\left(\frac{1}{f_{\omega}}\right)\left(\frac{m_{cp}}{M_p}\right)\left(\frac{r_{cp}}{R_p}\right)^{\frac{1}{2}}.
\label{q and l for disk}
\end{aligned}
\end{equation}
We obtained these expressions by imagining the disk to be composed of a set of infinitesimal rings of mass and integrating. 

To concretely examine typical values of $f_{\alpha}$, we must first define a set of fiducial system parameters. Here and in the remainder of the text unless otherwise noted, we use the parameters $k_2 = 0.125$ and $C = 0.25$ (close to the values for Uranus and Neptune, \citealt{1977Icar...32..443G, 2016CeMDA.126..145L}), $\rho_{\star} = 1.41 \ \mathrm{g/cm^3}$ (Solar density, which appears later), $\rho_p = 1.27 \ \mathrm{g/cm^3}$ (Uranus density), $\gamma = 3/4$ \citep{2002AJ....124.3404C},
and $f_{\omega} = \omega/\omega_b = 0.1$ \citep{2018NatAs...2..138B, 2018AJ....155..178B}. Additional parameters will be defined as necessary.

\begin{figure}
\epsscale{1.25}
\plotone{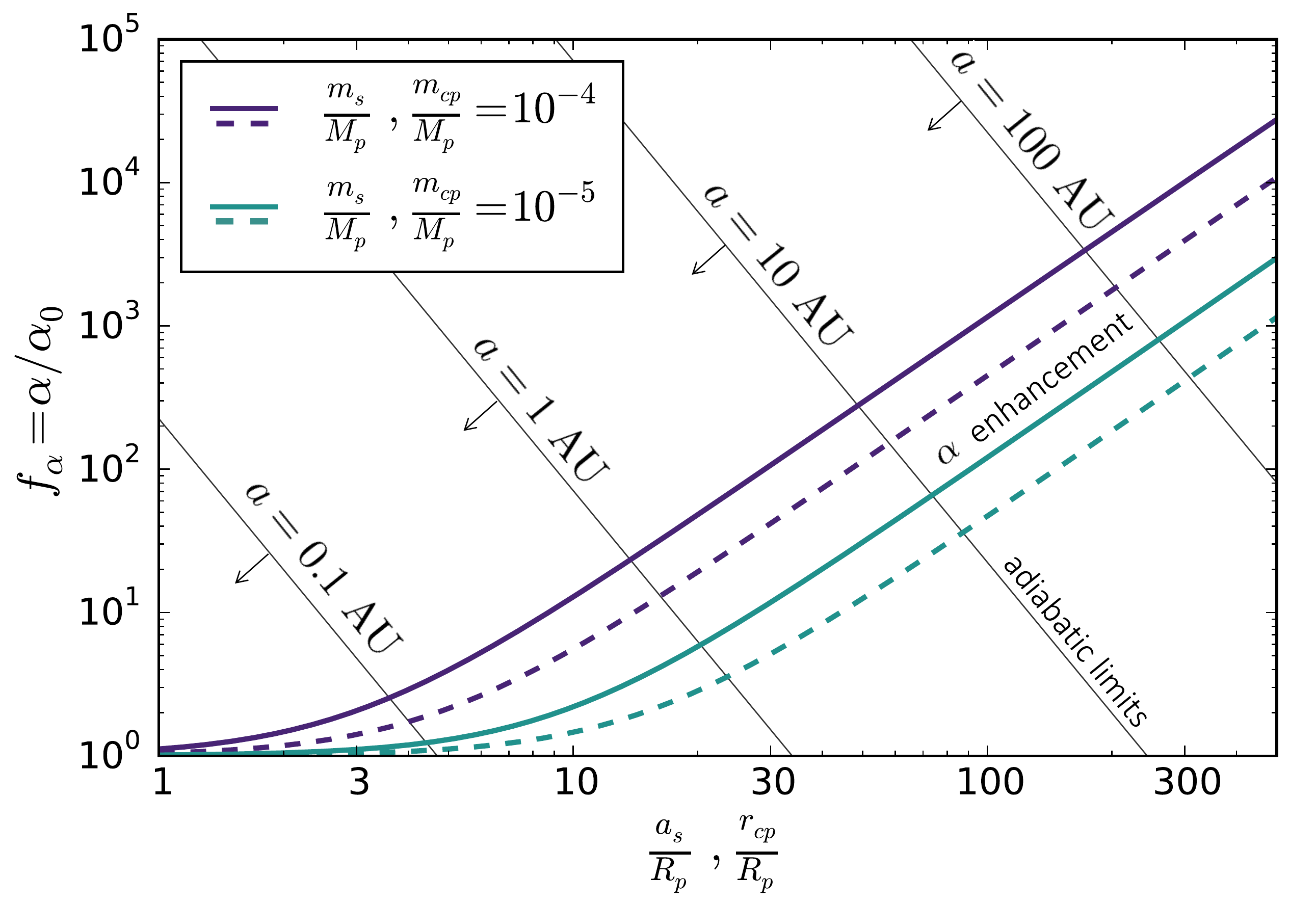}
\caption{Enhancement of the spin axis precession constant due to the presence of a single equatorial satellite (solid lines) or primordial circumplanetary disk (dashed lines). We plot the ratio $f_{\alpha} = \alpha/\alpha_0$ (where $\alpha_0$ is the precession constant in the satellite/disk-free case) as a function of $a_s/R_p$ or $r_{cp}/R_p$, the satellite's semi-major axis or disk's radius in units of planetary radii. The negative-sloped black lines show the adiabiatic limits in contours of planet semi-major axis. Below these lines, adiabatic gravitational coupling enables the system to precess as a unit.  
} 
\label{alpha_enhancement}
\end{figure}

Figure \ref{alpha_enhancement} shows the enhancement of the spin axis precessional constant due to a satellite (solid lines) or circumplanetary disk (dashed lines). It is clear to see that the enhancement is significant, reaching several orders of magnitude for large mass ratios and orbital distances. 

This $\alpha$-enhancement, however, is only active when the satellite/disk is close enough to the oblate planet to preserve adiabatic gravitational coupling. The adiabatic criterion requires that the rate of the satellite's nodal recession due to the planet's oblateness is much greater than the rate of the planet's spin axis precession \citep{1965AJ.....70....5G}. This criterion may be expressed
\begin{equation}
\lvert\nu_s\rvert \gg \alpha,
\end{equation}
where $\nu_s$ is the frequency of the satellite's nodal recession in the planet's equatorial plane \citep{1999ssd..book.....M},
\begin{equation}
\nu_s = \dot{\Omega}_s = -\frac{3}{2}n_s J_2\left(\frac{R_p}{a_s}\right)^2.
\end{equation}
Using $\alpha = f_{\alpha}\alpha_0$ and equation \ref{alpha0} for $\alpha_0$, the adiabatic criterion reduces to  
\begin{equation}
\frac{a_s}{R_p} \ll \left[\left(\frac{a}{R_{\star}}\right)^3
\frac{f_{\omega} C}{f_{\alpha}}\frac{\rho_p}{\rho_{\star}}\right]^{\frac{2}{7}},
\label{satellite coupling}
\end{equation}
where $\rho_p$ and $\rho_{\star}$ are the planet's and star's densities. 

Equation \ref{satellite coupling} specifies the maximum satellite semi-major axis for which the gravitational coupling will be upheld for a given $\alpha$-enhancement and planet semi-major axis. An equivalent formula exists for a circumplanetary disk, where $a_s$ is simply replaced by $r_{cp}$. The black diagonal lines in Figure \ref{alpha_enhancement} show contours of the planet semi-major axis calculated by equality of the left-hand and right-hand sides of equation \ref{satellite coupling}. At a given planet semi-major axis, satellite/disk orbital distances and $\alpha/\alpha_0$ enhancements below this line (toward the lower left) will maintain adiabatic gravitational coupling to the planet. 

\subsection{Orbit nodal recession}
In addition to the spin precessional motion of the planet and satellite/disk system, there are also torques that drive orbital precession. The protoplanetary disk's quadrupolar gravitational potential induces an orbit nodal recession for the planet with period, $T_g = 2\pi/g$. For a planet on a circular orbit, $g$ is defined \citep{2013ApJ...769...26C, 2013MNRAS.435..798T}
\begin{equation}
g = -\dot{\Omega} = \frac{3}{4}n\left(\frac{-\beta + 2}{1+\beta}\right)\left(\frac{1-\eta^{-1-\beta}}{1-\eta^{-\beta+2}}\right)\left(\frac{M_d}{M_{\star}}\right)\left(\frac{a}{R_o}\right)^3. 
\label{g}
\end{equation}
Here we have assumed the disk surface density profile follows a power law, 
\begin{equation}
\Sigma(a,t) = \Sigma_0(t)\left(\frac{a}{R_o}\right)^{-\beta},
\end{equation} 
where $R_o$ is the outer disk radius. The quantity $\eta = R_i/R_o$ is the ratio of inner and outer disk radii, and 
\begin{equation}
M_d(t) = \frac{2(1-\eta^{-\beta+2})\pi{R_o}^2}{-\beta+2}\Sigma_0(t)
\end{equation}
is the disk mass. We suppose that the disk profile follows the minimum-mass solar (or extrasolar) nebula \citep{1981PThPS..70...35H, 2013MNRAS.431.3444C} and set the power law exponent equal to $\beta=3/2$.

Early evolution of the system can cause the spin precession and disk-induced orbital precession to reach a commensurable rate. The resulting resonant encounter can excite the planetary obliquity to large values, sometimes up to $90^{\circ}$. We will explore the conditions for this resonant crossing and capture in the next section.

\section{Secular spin-orbit resonance}
\label{section 3}

Capture into a secular spin-orbit resonance occurs when $T_{\alpha}=2\pi/(\alpha\cos\epsilon_p)$ and $T_g=2\pi/g$ evolve such that $T_{\alpha}/T_g$ crosses through unity from above. As the protoplanetary disk decays, $\Sigma \rightarrow 0$ and $T_g \rightarrow \infty$. This crossing is therefore inevitable if $T_g < T_{\alpha}$ at some point during the system's evolution. The $T_g < T_{\alpha}$ crossing condition is equivalent to $g \gtrsim \alpha$ when the initial obliquities are small. Using equations \ref{alpha} and \ref{g} above and setting $\eta \ll 1$, the ratio $\alpha/g$ simplifies to 
%
\begin{equation}
\begin{split}
\frac{\alpha}{g} &= 0.004\bigg(\frac{k_2 f_{\omega}f_{\alpha}}{C}\bigg)\bigg(\frac{\rho_p}{\mathrm{\footnotesize{g/cm^3}}}\bigg)^{-\frac{1}{2}}\bigg(\frac{M_{\star}}{M_{\odot}}\bigg)^{\frac{1}{2}} \\
&\times\bigg(\frac{\eta}{0.001}\bigg)^{\frac{5}{2}}\bigg(\frac{M_d/M_{\star}}{0.01}\bigg)^{-1}\bigg(\frac{R_o}{100 \ \mathrm{AU}}\bigg)^3\bigg(\frac{a}{\mathrm{AU}}\bigg)^{-\frac{9}{2}}.
\end{split}
\end{equation}
By setting the initial value of $\alpha/g$ to be less than 1, we find the minimum semi-major axis at which the resonance will be crossed:
\begin{equation}
\begin{split}
a &> 0.3\ \mathrm{AU}\bigg[\bigg(\frac{k_2 f_{\omega}f_{\alpha}}{C}\bigg)\bigg(\frac{\rho_p}{\mathrm{\footnotesize{g/cm^3}}}\bigg)^{-\frac{1}{2}}\bigg(\frac{M_{\star}}{M_{\odot}}\bigg)^{\frac{1}{2}} \\
&\times\bigg(\frac{\eta}{0.001}\bigg)^{\frac{5}{2}}\bigg(\frac{M_d/M_{\star}}{0.01}\bigg)^{-1}\bigg(\frac{R_o}{100 \ \mathrm{AU}}\bigg)^3\bigg]^{\frac{2}{9}}.
\label{crossing criterion}
\end{split}
\end{equation}

Resonant crossing is a necessary but insufficient criterion for resonant capture. Capture into the spin-orbit resonance also requires that the crossing is adiabatic, in other words, that the crossing timescale is slow in comparison to the resonant libration period \citep{2004AJ....128.2510H}. This criterion may be expressed 
\begin{equation}
\dot{\alpha} - \dot{g} \lesssim \alpha g \sin\epsilon_0 \sin i,
\end{equation}
where $\epsilon_0$ is the planet's obliquity upon resonant crossing, and $i$ is the orbital inclination with respect to the invariable plane. The condition that $T_{\alpha}/T_g$ crosses unity from above (direction for capture) implies that $\dot{\alpha} - \dot{g} > 0$.

Disk evolution leads $M_d(t) \rightarrow 0$. Though this decay may not be described by a simply-defined function of time, if it is adiabatic, the sole relevant timescale is the decay rate upon resonant crossing, $\tau_{\scriptscriptstyle d}$. With this definition, $g/\dot{g} = M_d/\dot{M}_d = -\tau_{\scriptscriptstyle d}$. Unlike the decrease in $g$, which is clearly linked to the disk decay, there are a variety of mechanisms that could drive evolution in $\alpha$. These include Kelvin-Helmholtz contraction as the planet cools, dissipation of the circumplanetary disk, and disk-driven migration. We define $\tau_{\scriptscriptstyle \alpha} = -\alpha/\dot{\alpha}$ and the ratio $f_{\tau} = \tau_{\scriptscriptstyle d}/\tau_{\scriptscriptstyle \alpha}$. Finally, noting that $g \sim \alpha\cos\epsilon_0$ at resonant crossing and assuming $i$ is small, the capture criterion may be written
\begin{equation}
\frac{\alpha \ \tau_{\scriptscriptstyle d} \ i\sin2\epsilon_0}{2} \bigg(\frac{1}{\cos\epsilon_0-f_{\tau}}\bigg) > 1.
\end{equation}
For small $\epsilon_0$, this reduces to 
\begin{equation}
\alpha \ \tau_{\scriptscriptstyle d} \ i \epsilon_0 \bigg(\frac{1}{1-f_{\tau}}\bigg) > 1.
\end{equation}
This criterion highlights the fact that capture into a secular spin-orbit resonance cannot take place in a purely axi-symmetric system. Due to the infinite precession period at zero inclination, small but nevertheless finite inclination and obliquity are required for secular resonant coupling to ensue.

We substitute equation \ref{alpha} for $\alpha$ and solve for the maximum semi-major axis for which the crossing is adiabatic and capture is guaranteed,
\begin{equation}
\begin{split}
a &< 1.93\ \mathrm{AU}\bigg[\frac{k_2 f_{\omega}f_{\alpha}}{C}\bigg(\frac{\rho_p}{\mathrm{\footnotesize{g/cm^3}}}\bigg)^{-\frac{1}{2}}\bigg(\frac{1}{1-f_{\tau}}\bigg) \\
&\times\bigg(\frac{\tau_{\scriptscriptstyle d}}{\mathrm{Myr}}\bigg)\bigg(\frac{M_{\star}}{M_{\odot}}\bigg)\bigg(\frac{i}{1^{\circ}}\bigg)\bigg(\frac{\epsilon_0}{20^{\circ}}\bigg)\bigg]^{\frac{1}{3}}.
\label{capture criterion}
\end{split}
\end{equation}

\begin{figure} 
\epsscale{1.2}
\plotone{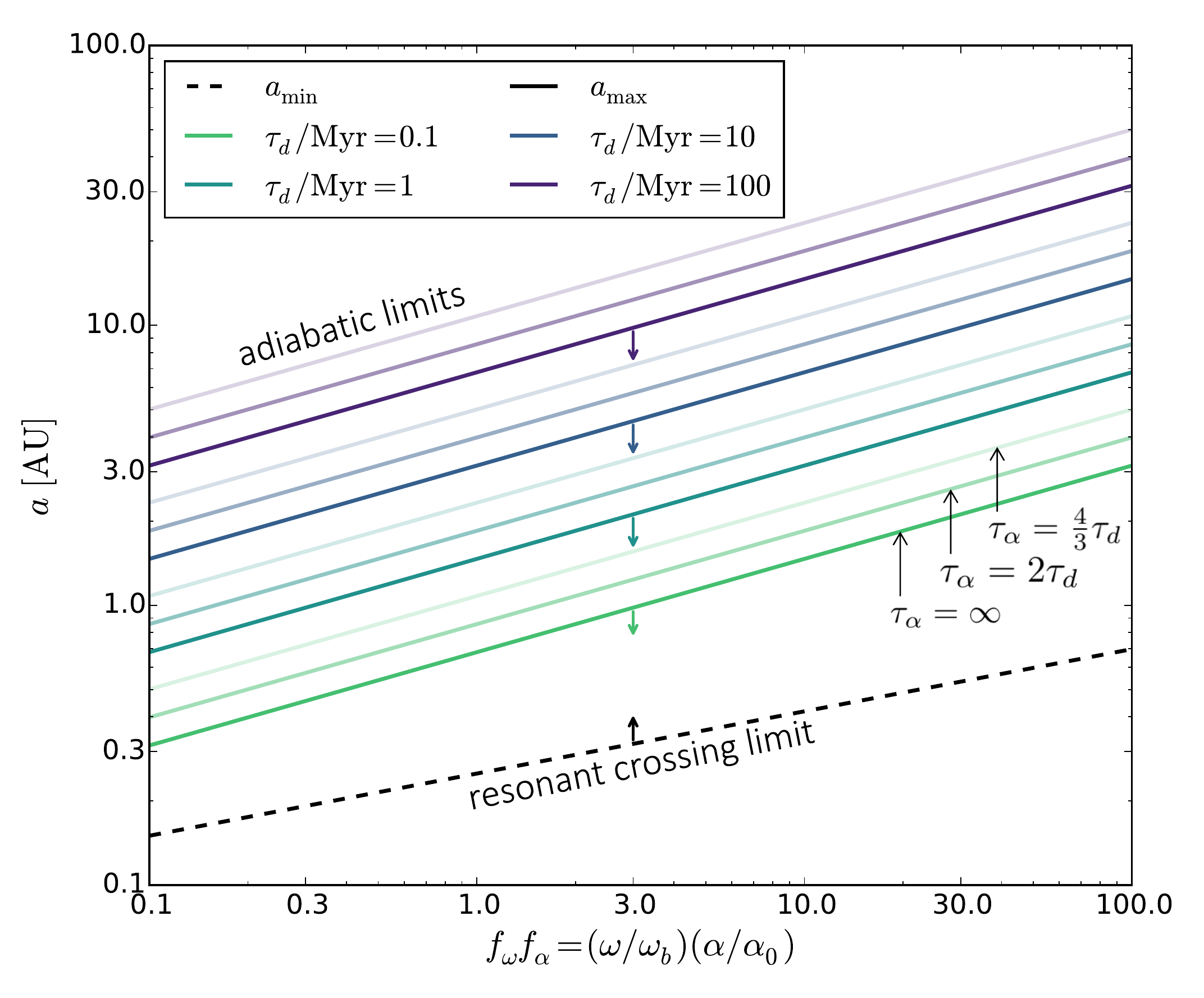}
\caption{The lower and upper limits of semi-major axis for which resonant crossing and capture will occur, plotted as a function of $f_{\omega}f_{\alpha} = (\omega/\omega_b)(\alpha/\alpha_0)$. The lower limit (dashed black line) is derived from the crossing criterion (equation \ref{crossing criterion}) and the upper limits (solid colored lines) from the adiabatic criterion (equation \ref{capture criterion}). In addition to system parameters defined in Section \ref{section 2}, these calculations assume that $i$, $\epsilon_0$, $M_{\star}$, $R_o$, $\eta$, and $M_d/M_{\star}$ all take on the fiducial values that they were presented with in the above equations. For each $\tau_d$, we show three values of $f_{\tau} = \tau_d/\tau_{\alpha}$ with different line transparencies.}
\label{resonant_crossing_capture}
\end{figure}

Figure \ref{resonant_crossing_capture} displays the lower and upper semi-major axis limits resulting from the resonant crossing (equation \ref{crossing criterion}) and capture (equation \ref{capture criterion}) criteria, respectively. For a given upper limit, the region above the line will result in resonant crossing but not capture because the adiabatic criterion is not upheld. (In that case the spin axis precession period is too long compared to the crossing timescale.) We plot these relations as a function of $f_{\omega}f_{\alpha} = (\omega/\omega_b)(\alpha/\alpha_0)$ for different crossing timescales. Notably, both $\tau_d$ and $\tau_{\alpha}$ affect the crossing timescale. The capture domain between the upper and lower limits widens as $\tau_{\alpha}$ approaches $\tau_d$ because this results in slower evolution of $g/{\alpha}$. The capture domain is infinite (no upper semi-major axis) in the limit $\tau_{\alpha} \rightarrow \tau_d$.

Beyond the upper limit in semi-major axis, the resonance will not be captured, but the crossing will still induce an obliquity increase due to an impulsive encounter with the resonant separatrix. The magnitude of this depends on the proximity of the crossing to the adiabatic limit. These non-adiabatic cases will be investigated in the following section.

\section{Analytical System Evolution}
\label{section 4}

\begin{figure*}
\epsscale{1.2}
\plotone{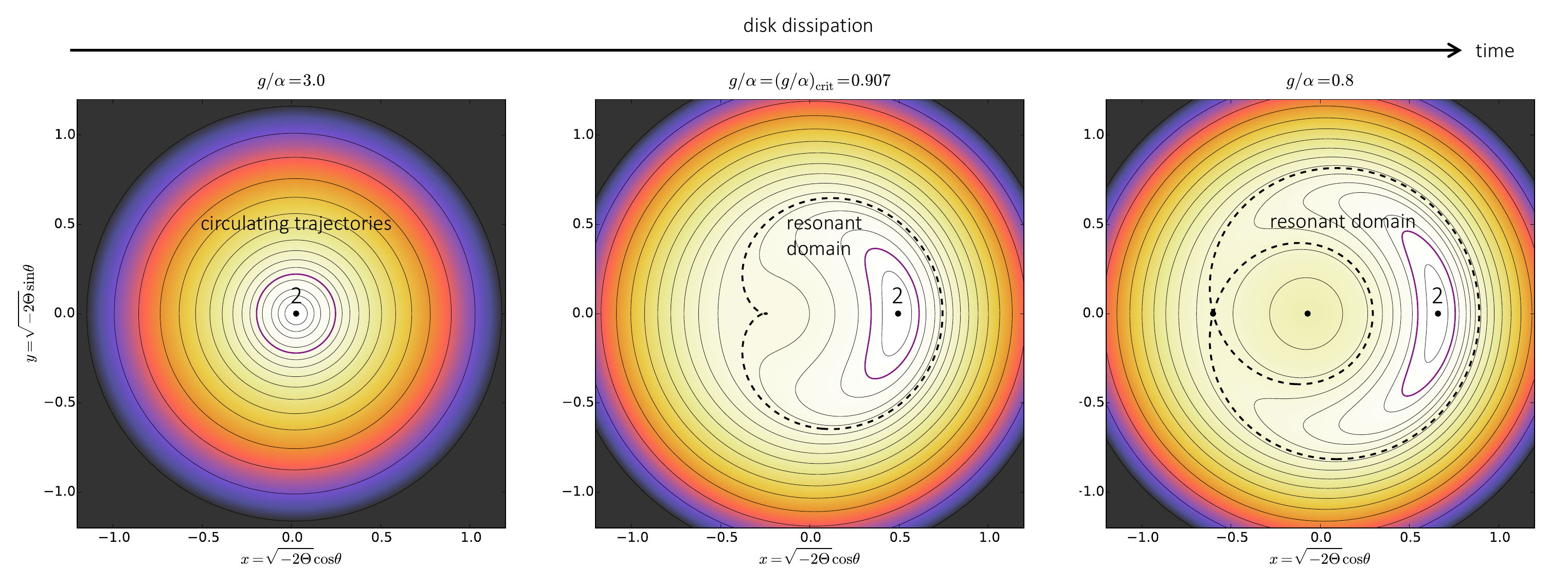}
\caption{Evolution of the phase space portrait of Hamiltonian \ref{Hamiltonian 4} as the protoplanetary disk dissipates, the ratio $g/\alpha$ decreases, and the spin-orbit resonance emerges. For all portraits, we use an inclination, $i = 1^{\circ}$, of the orbit off the disk plane, which determines the $(g/\alpha)_{\mathrm{crit}} = 0.907$ at which the separatrix appears. Recall that $\Theta = \cos{\epsilon} - 1$, so the polar radius, $\sqrt{-2\Theta}$, increases with the obliquity. The Cassini state 2 equilibrium is labeled with a ``2''. A trajectory that is bound to become resonant is accentuated in purple; the area enclosed by this trajectory is the same in the three panels.}
\label{phase space plot}
\end{figure*}

Here we consider an analytic perturbative model to describe the planet's spin state evolution upon encountering the secular resonance. The planet's spin evolution is best described in the non-inertial frame that precesses with the planet's orbit plane. We begin by defining canonical action-angle variables, $X=\cos\epsilon$ and $\psi$, where $\epsilon$ is the obliquity and $\psi$ is the angle between the projection of the spin vector onto the orbit plane and the ascending node. When the orbit nodal precession is uniform ($\dot{\Omega}$ is constant), the Hamiltonian that governs the system is given by \citep{2002mcma.book.....M}
\begin{equation}
\mathcal{H} = \frac{\alpha}{2}(1-e^2)^{-3/2}X^2 + \dot{\Omega}i\sqrt{1-X^2}\cos({\Omega + \psi}).
\label{Hamiltonian 1}
\end{equation}
Here, $\alpha$ is the spin axis precessional constant, $e$ is the eccentricity, and $i$ is the inclination of the orbit with respect to the invariable plane. The Hamiltonian of this system is well-studied; its origin, phase space structure, and equilibrium points have been characterized in many previous works \citep[e.g.][]{Colombo1966, Peale1969, 1974AJ.....79..722P, Ward1975, 1987CeMec..40..345H}.

To begin reducing the Hamiltonian to a time-independent, integrable form, we first apply a transformation of variables by introducing the canonically conjugate pair,
\begin{equation}
\Phi = 1-X \ \ \ \ \phi = -\psi.
\end{equation}
Hamiltonian \ref{Hamiltonian 1} now takes the form,
\begin{equation}
\begin{split}
\mathcal{H} &= \frac{\alpha}{2}(1-e^2)^{-3/2}(1-\Phi)^2 \\ &+ \dot{\Omega}i\sqrt{1-(1-\Phi)^2}\cos({\dot{\Omega}t - \phi}) + \mathcal{T}.
\label{Hamiltonian 2}
\end{split}
\end{equation}
Here we have also introduced a dummy action conjugate to time, $\mathcal{T}$, such that the Hamiltonian formally represents a two degree of freedom system.

Next, we remove the direct time-dependence by performing a canonical transformation arising from the following type-2 generating function: 
\begin{equation}
\mathcal{F}_2 = (\dot{\Omega}t - \phi)\Phi + (t)\Xi.
\end{equation}
The transformation equations yield the following new canonically conjugate variables \citep{2002mcma.book.....M}:
\begin{equation}
\begin{split}
\Theta &= -\Phi & \theta &= \dot{\Omega}t - \phi \\
\Xi &= \mathcal{T} - \dot{\Omega}\Theta & \xi &= t.
\end{split}
\end{equation}
It is clear from direct substitution of the above equations into Hamiltonian \ref{Hamiltonian 2} that the Hamiltonian is independent of $\xi$, such that $\Xi$ is a constant of motion and can be dropped. By applying the additional simplifications that $e=0$ and $g = -\dot{\Omega}$, we write the autonomous, one degree-of-freedom Hamiltonian in the form,
\begin{equation}
\mathcal{H} = \frac{\alpha}{2}(1+\Theta)^2 - g\Theta - g i\sqrt{1-(1+\Theta)^2}\cos{\theta}.
\label{Hamiltonian 3}
\end{equation}

An understanding of the dynamics encapsulated by Hamiltonian \ref{Hamiltonian 3} may be obtained by examining its level curves in phase space. The spin vector's motion is confined to reside on these level curves. Figure \ref{phase space plot} presents phase space portraits for three values of the ratio $g/{\alpha}$, which decreases as the disk decays. The equilibrium points (extrema of the Hamiltonian) are advected as $g/{\alpha}$ evolves. These equilibria are called ``Cassini states'' and are special configurations in which the spin vector remains fixed in the reference frame that precesses with the planet's orbit \citep{Colombo1966, Peale1969, 2015A&A...582A..69C}. At a critical value of $g/{\alpha}$ that depends on the orbital inclination \citep{2007ApJ...665..754F} and which we define as $(g/{\alpha})_{\mathrm{crit}}$, a separatrix emerges that partitions the phase space area and produces additional equilibria. (See middle panel of Figure \ref{phase space plot}.)

If the evolution of $g/{\alpha}$ is sufficiently slow compared to the dynamical timescale of the resonant motion, then adiabatic invariance dictates that the phase space area enclosed by a level curve trajectory is a conserved quantity, that is, as long as the orbit does not encounter a separatrix. Accordingly, consider a spin state that, at large $g/{\alpha}$, is prograde with a small obliquity. We illustrate this case with an accentuated purple trajectory in Figure \ref{phase space plot}. The spin vector initially circulates on a trajectory with a small phase space area. As $g/{\alpha}$ decreases and passes through $(g/{\alpha})_{\mathrm{crit}}$, the conservation of the phase space area implies that the spin vector enters the resonant domain and starts librating about the equilibrium on a small, banana-shaped level curve. This resonant equilibrium is ``Cassini state 2'', and it corresponds to a configuration in which the spin vector and orbital angular momentum vector \textit{coprecess} on opposite sides of the vector normal to the invariable plane. Following resonant capture, the planet's obliquity is forced to larger and larger values as $g/{\alpha}$ decreases. Specifically, the condition $g \approx \alpha\cos\epsilon$ is upheld when the orbital inclination is small. Moreover, any source of dissipation, such as tides, will tend to force the system toward the equilibria.

\vfill\null

\subsection{Outcome of resonant encounter}

\begin{figure}
\epsscale{1.13}
\plotone{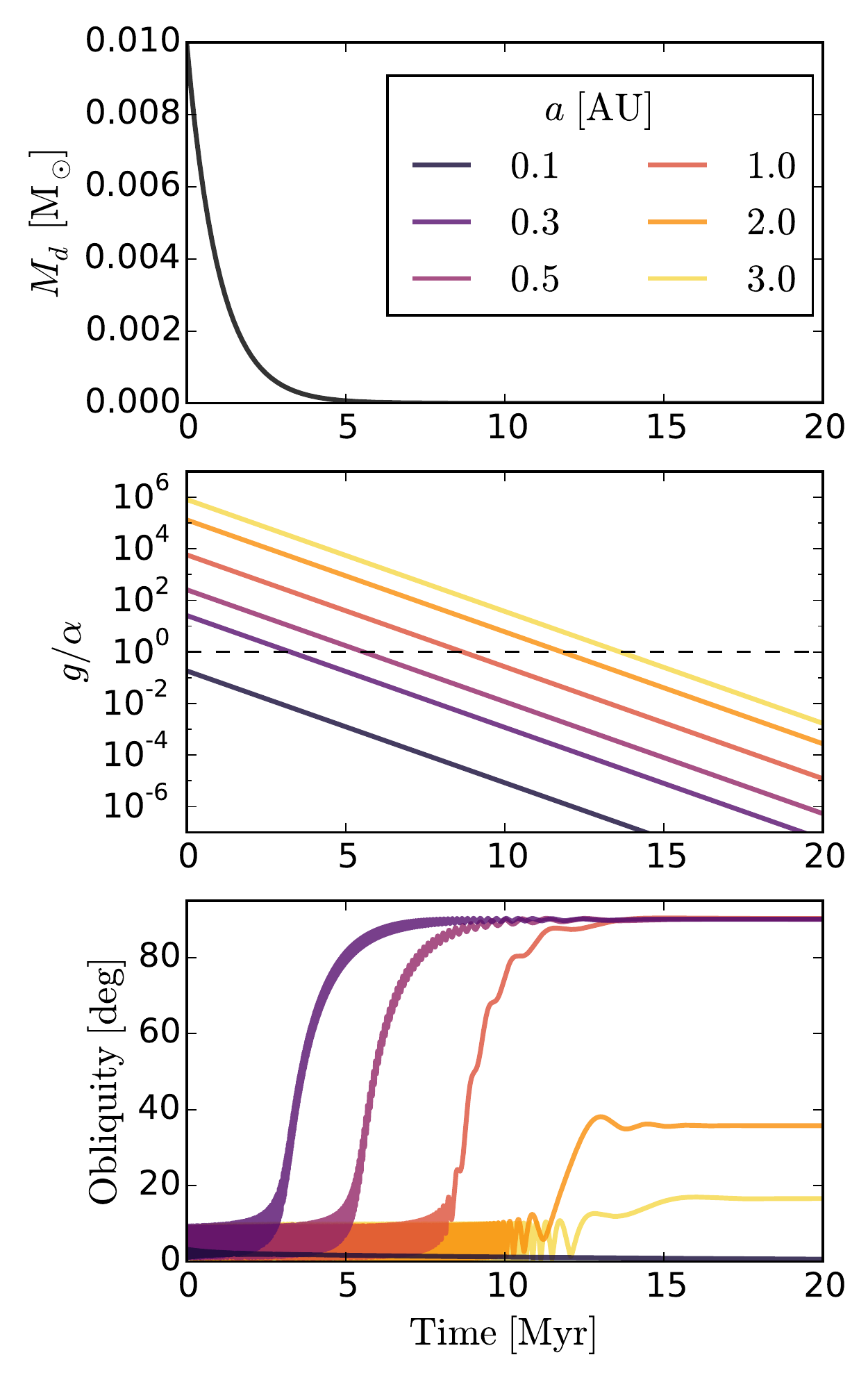}
\caption{Obliquity evolution of example systems obtained via integration of Hamiltonian \ref{Hamiltonian 4}. All simulations use $\tau_d = 1 \ \mathrm{Myr}$ and the same disk mass evolution but have different planet semi-major axes. \textit{Top panel}: Protoplanetary disk mass as a function of time. \textit{Middle panel}: The ratio of orbital and spin axis precession frequencies. \textit{Bottom panel}: Planetary obliquity evolution.  }
\label{example_simulations}
\end{figure}

With a perturbative Hamiltonian now in hand, we use it to efficiently explore the outcome of resonant encounter for different system configurations. Formally speaking, Hamiltonian \ref{Hamiltonian 3} is integrable only if $\alpha$ and $g$ are constant. If, however, $\alpha$ and $g$ evolve much slower than the libration period of $\theta$, then successive phase-space portraits entailed by integrable Hamiltonian \ref{Hamiltonian 3} provide an excellent approximation to the real dynamics. Thus, it is instructive to solve the evolution of the system by an application of Hamilton's equations. We do so numerically but first apply a final canonical transformation to convenient Cartesian coordinates,
\begin{equation}
x = \sqrt{-2\Theta}\cos\theta \ \ \ \ 
y = \sqrt{-2\Theta}\sin\theta.
\end{equation}
The Hamiltonian in these coordinates is
\begin{equation}
\mathcal{H} = \frac{\alpha}{2}\bigg(1-\frac{1}{2}x^2 - \frac{1}{2}y^2\bigg)^2 + \frac{1}{2}g(x^2+y^2) - \frac{i}{2}gx\sqrt{4-x^2-y^2}.
\label{Hamiltonian 4}
\end{equation}

There are two principal parameters that govern whether the secular resonance will be crossed and captured: the semi-major axis, $a$, and disk decay timescale, $\tau_{\scriptscriptstyle d}$. We reduce the parameter space of potential system configurations to these two parameters and fix all others to the fiducial values assigned in Section \ref{section 2} as well as the following: $M_{\star} = M_{\odot}$, $M_d(t=0) = 0.01 \ M_{\sun}$, $R_o = 100 \ \mathrm{AU}$, $\eta = 0.001$, $i = 5^{\circ}$, $f_{\alpha} = 1$, and $f_{\omega} = 0.1$ \citep{2018NatAs...2..138B, 2018AJ....155..178B}. 

We keep the spin axis precession rate fixed ($\alpha = \mathrm{const}$, $f_{\tau} = 0$), such that all evolution of the system is due to disk dissipation. The protoplanetary disk mass is set to evolve as a simple exponential decay, 
\begin{equation}
M_d(a,t) = M_d(t=0)\exp(-t/\tau_{\scriptscriptstyle d}).
\end{equation}
While this is certainly a simplification of the system's evolution, it is sufficient for providing a sense of the behavior, which only depends on the evolution of the ratio $g/{\alpha}$. We note that in general, the spin rate, planetary radius, and possibly semi-major axis can also change during this time. Ultimately, however, $g/{\alpha}$ will cross through unity from above, and this is encapsulated by our model. With this parameterization in place, we apply Hamilton's equations to Hamiltonian \ref{Hamiltonian 4} and integrate numerically with a conventional Runge-Kutta ODE solver.

Five example system evolutions are displayed in Figure \ref{example_simulations}. They use the same parameters except for the semi-major axes. As shown in the second panel, all examples except for the $a = 0.1$ AU case sweep through the resonance at some point during the disk decay. For the $a = 0.3$, $0.5$, and $1.0$ AU cases, the resonance crossing is adiabatic and the planetary obliquities get excited all the way to $90^{\circ}$. The resonance crossings in the $2$ AU and $3$ AU cases are too fast for capture, so the planetary obliquities only receive impulsive, limited excitations. 

\begin{figure}
\epsscale{1.25}
\plotone{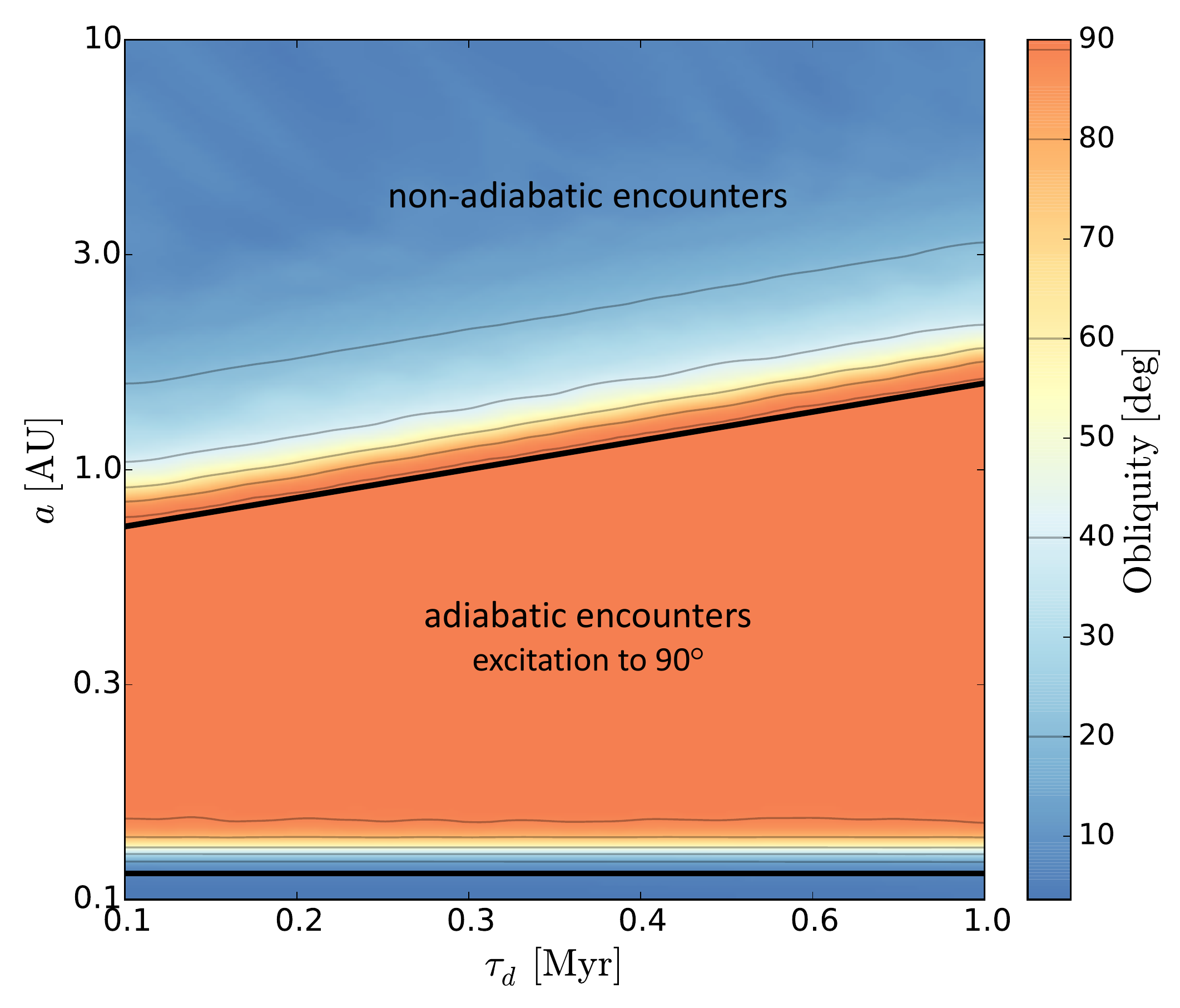}
\caption{A map of the planet's final obliquity following passage through the secular resonance at different semi-major axes and disk dissipation timescales. The colors and thin contours (whose values correspond to the thin lines in the colorbar) show the results using integrations with Hamiltonian \ref{Hamiltonian 4}. The thick black lines show the analytic lower and upper semi-major axis limits. The lower limit is the critical semi-major axis beyond which resonant crossing exists (equation \ref{crossing criterion}), and the upper limit is the critical value below which there is guaranteed resonant capture that excites the obliquity to $90^{\circ}$ (equation \ref{capture criterion}). 
} 
\label{final_obliquities_heatmap}
\end{figure}

Individual examples like these are helpful, but it is more informative to summarize the obliquity evolution over a wide range of system parameters. Figure \ref{final_obliquities_heatmap} shows the final obliquity following resonant encounters for a grid of $a$ and $\tau_{\scriptscriptstyle d}$. The results are in excellent agreement with the analytic critical semi-major axis limits for resonant crossing and capture that we derived with equations \ref{crossing criterion} and \ref{capture criterion}. Below the upper limit, the obliquity is excited all the way to $90^{\circ}$ if the resonance is crossed. The heatmap also shows the magnitude of the obliquity excitation that occurs when the resonance is crossed too quickly for capture. This information is difficult to access through analytic considerations alone. Though it is possible to excite the obliquity without resonant capture, the excitation is not particularly significant unless it is relatively close to the adiabatic limit. However, excitations to $\sim10-20^{\circ}$ still occur far past the limit. 

\subsection{Limiting effects from perturbing planets}
\label{subsection 4.2}
In our model thus far, we have considered the interactions between a single planet, its host star and the protoplanetary disk in which it forms. In actuality, planets usually do not form in isolation, but rather in the presence of companion planets. Here we examine how the disk-induced secular spin-orbit resonance mechanism is altered as it pertains to multiple-planet systems.

For definitiveness, consider the simplified example where the perturbing planet's orbit is taken to coincide with the plane of the disk. In this $i'=0$ case, the frequency of a planet's nodal recession, $g = -\dot{\Omega}$, never reaches zero. Rather, as the disk density diminishes, $g$ reaches a floor set by the secular frequencies of nodal recession induced by planet-planet perturbations. In the low-$i$, low-$e$ limit, these frequencies may be calculated using Laplace-Lagrange secular theory \citep{1999ssd..book.....M}. Specifically, the rate of nodal regression forced by a planetary companion reads:
\begin{equation}
g_{pp} = \frac{n}{4} b_{3/2}^{(1)}(\alpha)\alpha\bar{\alpha}\frac{M^{\prime}}{M_{\star}}
\label{LL g}
\end{equation} 
where $\alpha = a/a^{\prime}$, $\bar{\alpha} = \alpha$ if the perturbation is external and is unity otherwise, while the constant, $b_{3/2}^{(1)}(\alpha)$, is a Laplace coefficient defined by 
\begin{equation}
b_{3/2}^{(1)}(\alpha) = \frac{1}{\pi}\int_{0}^{2\pi}\frac{\cos\psi}{(1-2\alpha\cos\psi + \alpha^2)^{3/2}}d\psi.
\end{equation}
Although beyond the scope of our discussion, we note that when $i' \ne 0$, the scenario is marginally more complicated due to multiple frequency components that are present (those associated with both the planet-disk interactions and the planet-planet interactions), implying that obliquity excitation in that case is not necessarily limited.

The floor on the nodal recession frequency can in some cases prevent resonant obliquity excitation, depending on the comparison between the planet's spin axis precession frequency and the secular frequencies of nodal recession. If $g_{pp} > \alpha$, the nodal recession never reaches a commensurability with the spin axis precession, and obliquity excitation is blocked. Alternatively, if $g_{pp} < \alpha$, then limited excitation up to $\epsilon \approx \cos^{-1}(g_{pp}/{\alpha})$ will occur before the floor is reached. As will be shown below, it is the suppression of the outlined obliquity excitation mechanism by planet-planet interactions that renders this process inapplicable to the Solar System's ice giants.

\section{Application to Uranus and Neptune}
\label{section 5}

Our analysis has indicated that disk-induced secular spin-orbit resonance is less likely for planets at distances greater than $\sim 5$ AU. This can be seen from the $a \sim 2$ AU fiducial upper limit at which the crossing is adiabatic and capture is ensured (see for example equation \ref{capture criterion}). Are there conditions for which the mechanism is relevant beyond $a \gtrsim 10$ AU? Figure \ref{resonant_crossing_capture} illustrates that planets with large $a$ are affected in the limit of significantly enhanced $\alpha$ (large $f_{\alpha}$), slow protoplanetary disk dissipation timescale $\tau_d$, and $f_{\tau} = \tau_{d}/{\tau_{\alpha}}$ close to unity. With all of these factors working together, the mechanism might be relevant to the ice giants, Uranus and Neptune. 

The origins of Uranus's $98^{\circ}$ and Neptune's $30^{\circ}$ obliquities are not known with certainty. Their spin precession rates are much slower than any secular eigenfrequencies \citep{2006Icar..185..312B, 2018ARA&A..56..137N}, implying that secular spin-orbit resonances induced by planet-planet interactions are not the cause of their large obliquities unless the spin precession was significantly enhanced by a massive satellite or circumplanetary disk \citep{2010ApJ...712L..44B}, or from the planets originally being closer to the Sun \citep{2018DDA....49P..10R}. 

Collisional tilting from roughly Earth-mass giant impacts is the most well-accepted theory for the tilt of Uranus and Neptune \citep{1966SvA.....9..987S, 1990Icar...84..528K, 1992Icar...99..167S, 2012Icar..219..737M, 2018ApJ...861...52K}. To explain Uranus's prograde equatorial satellite system, \cite{2012Icar..219..737M} showed that the obliquity prior to the last tilting event must have been non-negligible, leading them to suggest that a multitude of such impacts were required. A particularly strong impact would evaporate ice from the ejected debris \citep{2004A&A...413..373M}, making the satellites devoid of ice, in contradiction with their $\sim 50\%$ water ice abundance \citep{2018ApJ...868L..13S}. Apart from collisions, the obliquities of Uranus and Neptune may also have been altered by some early process that twisted the Solar System's total angular momentum vector \citep{1991Icar...89...85T}.

Disk-induced secular spin-orbit resonance is yet another option to consider. Here we outline the idealized scenario under which it could have been relevant. First, resonant capture would have required that $\alpha$ was enhanced by the presence of a circumplanetary disk. \cite{2018ApJ...868L..13S} used three-dimensional radiative hydrodynamic simulations to show that the ice giants likely did form gaseous circumplanetary disks during their accretion. The disk masses in their simulations were $m_{cp} \sim 10^{-3} M_p$. Moreover, the disks were capable of forming icy satellites matching the current-day moon system of Uranus, which itself matches the $\sim 10^{-4}M_{p}$ mass ratio of Jupiter's and Saturn's satellite systems \citep{2006Natur.441..834C}. We therefore assume the existence of a circumplanetary disk in the calculations that follow in section \ref{section 5.1}. \\

\subsection{Idealized conditions for capture}
\label{section 5.1}

\begin{figure}
\epsscale{1.25}
\plotone{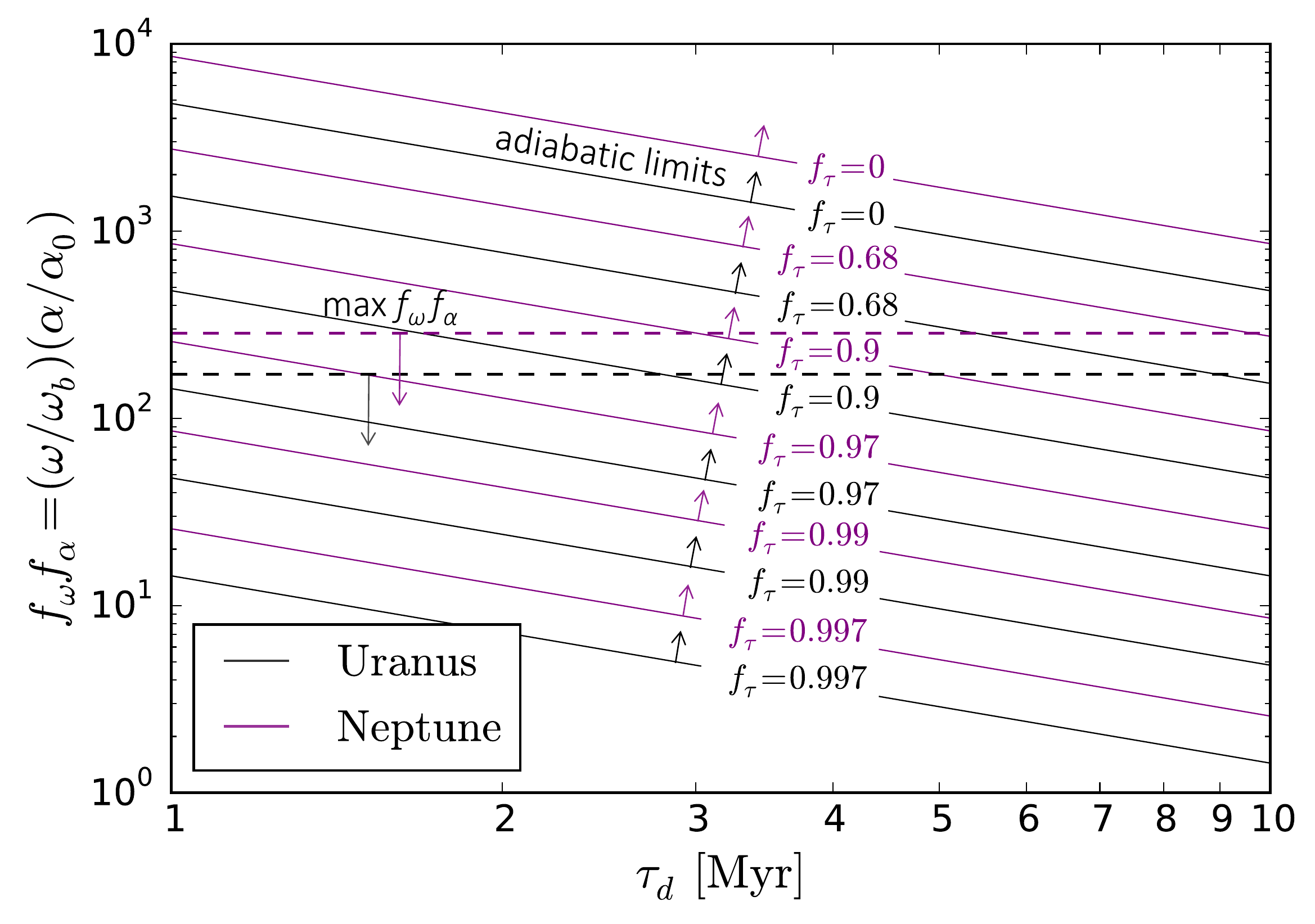}
\caption{Resonant capture domain for Uranus and Neptune in $\tau_d$ and $f_{\omega}f_{\alpha} = (\omega/\omega_b)(\alpha/\alpha_0)$ parameter space. For a given $f_{\tau} = \tau_d/{\tau_{\alpha}}$, the region above the line guarantees resonant capture through the adiabatic criterion (equation \ref{capture criterion}). The horizontal dashed lines mark the maximal $f_{\omega}f_{\alpha}$ assuming that a circumplanetary disk is providing the $\alpha$ enhancement.} 
\label{Uranus_Neptune_adiabaticity_contours}
\end{figure}

We begin with the condition \ref{capture criterion} for resonant capture determined by the adiabatic criterion. Using known planetary parameters for Uranus and Neptune, we can use this to find the values of $f_{\omega}$, $f_{\alpha}$, $f_{\tau}$, and $\tau_d$ required for resonant capture. We use present-day values of the semi-major axes and densities, Love numbers equal to $k_{2U} = 0.104$ and $k_{2N} = 0.127$ \citep{2016CeMDA.126..145L}, and moment of inertia factors $C_U = C_N = 0.225$. It is likely that these parameters were different early on, but because we have little constraint on their histories, it is best to use present-day values for this plausibility argument. Finally, we use inclinations, $i_U = 1^{\circ}$ and $i_N = 2^{\circ}$. 

Figure \ref{Uranus_Neptune_adiabaticity_contours} shows the resulting limits on the resonant capture domain. For a line with a given $f_{\tau}$, the space above the line represents the $f_{\omega}f_{\alpha}$ and $\tau_d$ required to guarantee capture by the adiabatic constraint. We also plot the maximum values of $f_{\omega}f_{\alpha}$ by assuming $f_{\omega} = 1$, considering a circumplanetary disk with $m_{cp}/M_p = 10^{-3}$, and applying equations \ref{alpha0}, \ref{alpha}, and \ref{q and l for disk} to calculate $f_{\alpha}$. We use the largest $r_{cp}/R_p$ that is still below the limit for adiabatic gravitational coupling (equation \ref{satellite coupling}).

Clearly, capture is difficult to attain. The resonant domain corresponds to large $f_{\omega}f_{\alpha}$, implying significant $\alpha$-enhancement from a circumplanetary disk or satellites. It also requires long timescales for $\tau_d$ unless $f_{\tau}$ is nearly unity. $f_{\tau} \sim 1$ can only occur if $\alpha$ is evolving at a similar rate as $g$. It turns out, however, that this is a good assumption, which we will now show.

Early on in the system's evolution, $\rho_p$, $k_2$, $C$, and $f_{\omega}$ all evolve non-trivially, but $m_{cp}$ changes the most. Accordingly, let us consider the simplification that $m_{cp}$ is the \textit{only} changing quantity. Then $\tau_{\alpha}$ becomes
\begin{equation}
\tau_{\alpha} = -\frac{\alpha}{\dot{\alpha}} = -\left[\frac{\dot{q}}{(k_2 {f_{\omega}}^2/3)+q} - \frac{\dot{l}}{C+l}\right]^{-1}.
\label{tau_alpha}
\end{equation}
Equations \ref{q and l for disk} show that $q/{\dot{q}} = l/{\dot{l}} = m_{cp}/{\dot{m}_{cp}}$. 
If $r_{cp}/R_p \sim 100$, we can take the limits $q \gg k_2 {f_{\omega}}^2/3$ and $l  \ll C$. Equation \ref{tau_alpha} is then approximately 
\begin{equation}
\tau_{\alpha} \approx -\frac{q}{\dot{q}} = -\frac{m_{cp}}{\dot{m}_{cp}}.
\end{equation}
We simplify further by assuming that $m_{cp} \approxprop M_d$. While this approximation is theoretically-motivated as a result of the scaling between the circumstellar and circumplanetary disk surface densities \citep{2010AJ....140.1168W, 2018ApJ...868L..13S}, in reality both $m_{cp}$ and $M_d$ evolve non-trivially. The $m_{cp} \approxprop M_d$ simplification implies however that $m_{cp}/\dot{m}_{cp} \approx M_d/\dot{M}_{d}$ and finally that $\tau_{\alpha} \approx \tau_d$ and $f_{\tau} \approx 1$. Therefore, this first-order argument suggests that it is reasonable that $f_{\tau}$ is close to unity, which greatly alleviates the constraints posed by Figure \ref{Uranus_Neptune_adiabaticity_contours}. 

\begin{figure}
\epsscale{1.25}
\plotone{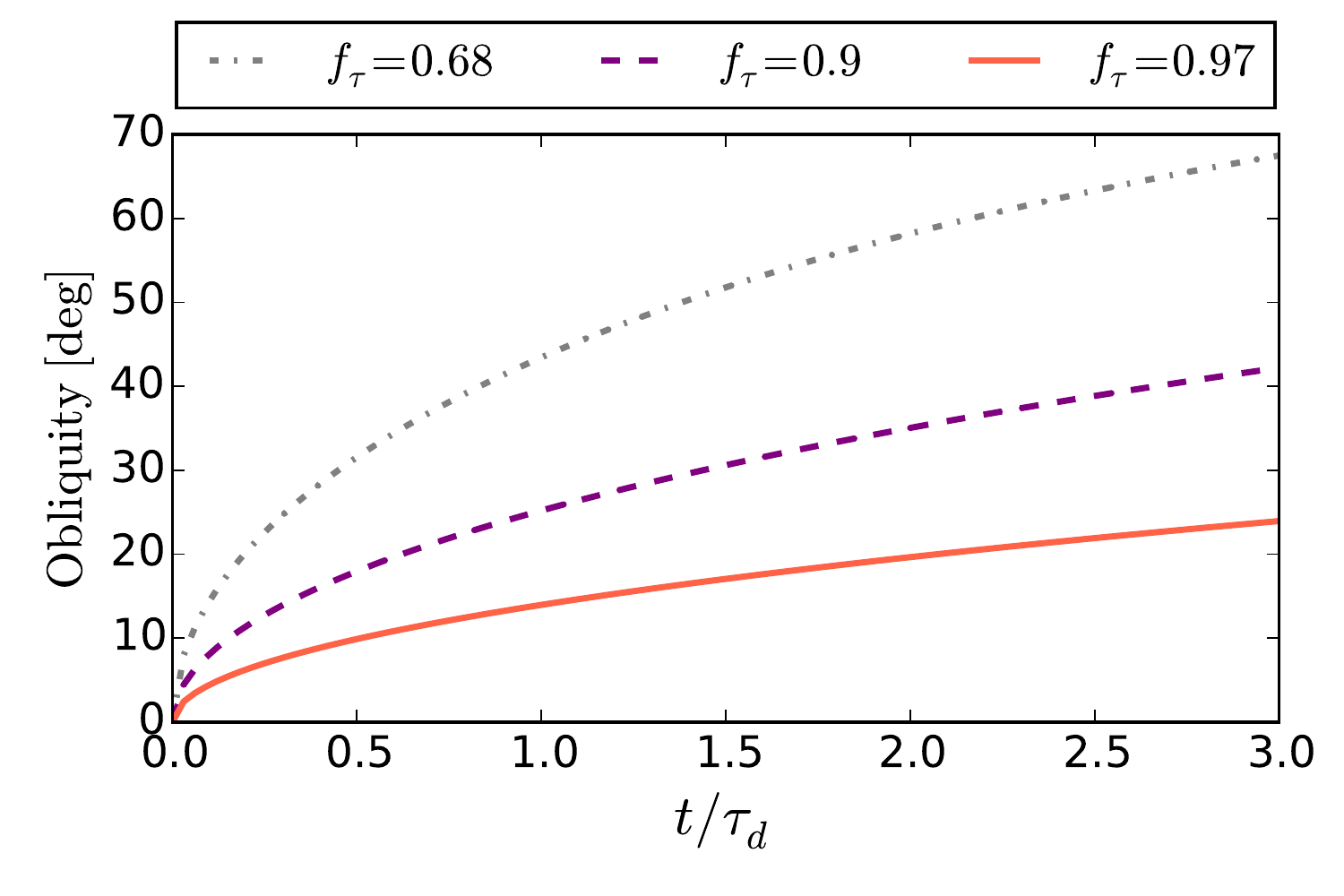}
\caption{Obliquity as a function of time (normalized by $\tau_d$) after resonant capture. We show the evolution for different $f_{\tau} = \tau_d/\tau_{\alpha}$, corresponding to three of the values shown in Figure \ref{Uranus_Neptune_adiabaticity_contours}. We use $\epsilon_0 = 0$ for simplicity, but the initial obliquity must be at least slightly non-zero. The disk lifetime is expected to be a few times $\tau_d$.} 
\label{obliquity_vs_time}
\end{figure}

In addition to the adiabatic constraint, another limitation is the length of time required to tilt the planet while the protoplanetary disk is still present. The resonant tilting scenario is only plausible if it can be completed sufficiently quickly. Once the resonance is captured, $g \approx \alpha\cos\epsilon$ holds if the inclination is small, and the obliquity evolves as 
\begin{equation}
\epsilon(t) = \cos^{-1}\left[\cos\epsilon_0 \exp\left(-(1-f_{\tau})\frac{t}{\tau_d}\right)\right],
\end{equation}
where we've assumed constant $\tau_d$ and $f_{\tau}$. Figure \ref{obliquity_vs_time} shows $\epsilon(t)$ as a function of $t/{\tau_d}$ for several values of $f_{\tau}$. Obliquity excitations on the order of $\epsilon \sim 10^{\circ} - 30^{\circ}$ -- which notably includes the obliquity of Neptune -- are easily attainable within the disk lifetime (a few times $\tau_d$) for a range of $f_{\tau}$'s. On the other hand, tilting a planet from near $0^{\circ}$ to near $90^{\circ}$ most likely cannot be accomplished while the disk is still present. This timescale constraint is also prohibitive in the mean-motion/spin precession resonant tilting scenario for Uranus \citep{2018CeMDA.130...11Q}. The tilting timescale might be shortened, however, if $\tau_d$ and $f_{\tau}$ are not constant. 

\subsection{Suppression due to giant planet perturbations}

The preceding discussion has outlined the fine-tuned conditions required for disk-induced obliquity excitation to be relevant for Uranus and Neptune. It is likely, however, that this scenario as envisioned cannot have occurred due to the suppression mechanism we discussed in section \ref{subsection 4.2}, which in this case involves secular interactions among the giant planets.

In the $i'=0$ case where the perturbing planets (i.e. Jupiter and Saturn) are aligned with the disk, Uranus's and Neptune's nodal recession frequencies would have decreased during disk dispersal but reached a floor at the secular frequencies associated with interactions among the four giant planets. Obliquity excitation would not have been possible because the secular frequencies -- shown in Table \ref{SS secular frequencies} -- are faster than the spin axis precession constants of Uranus and Neptune. Specifically, when $\omega = \omega_b$ (as opposed to the present-day spin rates), $\alpha_0 = 0.014$ arcsec yr$^{-1}$ for Uranus $\alpha_0 = 0.0040$ arcsec yr$^{-1}$ for Neptune. Even when enhanced by circumplanetary disks up to a factor of 100-1000, the spin axis precession rates are not faster than the fastest secular frequency. Therefore, if $i' = 0$, the resonance capture mechanism would have been inhibited.

\begin{table}[h]
\centering
\caption{Nodal secular frequencies for the giant planets in the Solar System \citep{1988A&A...198..341L}. In the notation of \cite{1999ssd..book.....M}, the subscripts $i$ are the eigenmode numbers, and $f_i$ are the eigenfrequencies associated with the inclination/node solution. Planet $i$ is dominated by the mode of the same number.}
\begin{tabular}{c | c}
$i$ & $f_i$ (arcsec yr$^{-1}$)   \\
\hline
5 & -    \\
6 & -26.34 \\
7 & -2.99 \\
8 & -0.69
\end{tabular}
\label{SS secular frequencies}
\end{table}


\section{Summary and Discussion}
\label{section 6}

Secular spin-orbit resonance is a robust mechanism for producing non-zero planetary axial tilts. While previous work has mainly focused on resonances involving orbital precession induced by planet-planet interactions, here we addressed primordial orbital precession driven by the gravitational field of a protoplanetary disk. Unlike planet-planet interactions, disk-driven orbital precession is transient. However, a wide range of precession frequencies are swept during the disk decay process, thereby producing many opportunities for obliquity-exciting commensurabilities between the planet's spin and orbital precession frequencies.

We find that planets with semi-major axes, $0.3 \ \mathrm{AU} \lesssim a \lesssim 2 \ \mathrm{AU}$, can readily be tipped to $90^{\circ}$ if the interaction is adiabatic. The details of these limits depend on the exact parameter configuration, and the upper limit on $a$ increases significantly (up to $\sim 20 \ \mathrm{AU}$) if the disk dissipates slowly or if the planet is accompanied by an evolving circumplanetary disk or massive satellite(s). Moreover, beyond the upper limit in $a$, the resonance will still be encountered impulsively and will excite the obliquity to a degree that depends on the proximity to adiabaticity. We employed a perturbative Hamiltonian model to show that the obliquity may be excited up to $\sim 10^{\circ}-20^{\circ}$ even far past the adiabatic limit. However, planet-planet perturbations can in some cases prevent obliquity excitation by providing a floor on the nodal recession frequency.

Uranus's $98^{\circ}$ and Neptune's $30^{\circ}$ obliquities are typically ascribed to giant impacts. Smooth particle hydrodynamic simulations show that a single impactor on Uranus would necessarily have been quite massive ($\gtrsim 2 \ M_{\oplus}$, \citealt{2018ApJ...861...52K, 2019arXiv190109934K}). If it was too large, however, the impact would have evaporated ices from the ejected debris \citep{2004A&A...413..373M, 2018ApJ...868L..13S}, which presents tension with the icy composition of Uranus's prograde equatorial satellites. Moreover, Uranus's satellite configuration could have only formed if the pre-impact obliquity was non-negligible, comparable with that of Neptune's present-day $30^{\circ}$ tilt \citep{2012Icar..219..737M}. This suggests an additional impact or some other tilting mechanism. 

Disk-induced secular spin-orbit resonance may present a viable alternative to collisional tilting in an idealized setting. Uranus and Neptune likely formed gaseous circumplanetary disks during their formation \citep{2018ApJ...868L..13S}. These disks would have increased the rate of spin axis precession and caused the spin and orbital precession frequencies to evolve on similar timescales, allowing a potential resonant encounter to last far longer and the possibility of obliquity excitation up to $\sim30^{\circ}$. This excitation would have been inhibited, however, due to secular interactions with Jupiter and Saturn if they were aligned with the disk.


If disk-induced spin-orbit resonance might have played a role in Uranus's and/or Neptune's elevated obliquities, the question is raised for the rest of the Solar System planets: were they affected? One important consideration is that the mechanism requires non-zero orbital inclinations. The inclinations of Jupiter and Saturn at the time of resonance crossing may have been much smaller than those of Uranus and Neptune, such that adiabatic excitation was not a possibility. Moreover, the primordial obliquities of the gas giants would have been perturbed after disk dispersal by their planet interaction-induced spin-orbit resonances \citep{Ward2004, 2004AJ....128.2510H, 2006ApJ...640L..91W}. As for the terrestrial planets, they likely finished forming through collisional growth after nebular gas dispersal (see \citealt{2018haex.bookE.142I} for a review), and such collisions -- as well as chaotic obliquity variations arising from planet interactions \citep{1993Natur.361..608L} -- would have altered their primordial obliquities.

Extrasolar planets are prevalent in the $0.3 \ \mathrm{AU} \lesssim a \lesssim 2 \ \mathrm{AU}$ range where the mechanism is most relevant. We therefore predict that it is important for many exoplanets, although their excited obliquities might be perturbed by subsequent giant impacts. Close-in planets on the short end of the semi-major axis range will also be susceptible to tidal interactions that gradually right any planetary tilt produced by the resonant encounter. During this damping process, however, spin-orbit resonances induced by planet-planet interactions can be captured. The disk-induced resonance thus plays an important role for close-in systems by increasing the likelihood that the planets later find a separate resonance that, even in the presence of tides, captures the obliquity in a long-term excited state.

A similar staging role can be played by spin-orbit resonance induced by the oblateness of a rapidly-rotating young star \citep{2019NatAs.tmp..218M}. Apart from the disk, early on in the system's lifetime there is an additional component to each planet's orbit nodal recession from the quadrupole field of the young star \citep{2013ApJ...778..169B, 2017AJ....154...93S}. Like the disk decay, this can produce obliquity-exciting resonant encounters as the star contracts and spins down. It is most relevant for close-in planets, since the nodal recession frequency is a steep inverse function of $a$. 

The disk-induced secular spin-orbit resonance is a powerful mechanism for exciting non-zero planetary obliquities with many areas ripe for further exploration. For instance, it would be useful to investigate how warps and turbulence in the disk affects the first-order picture we have developed here. Gaps and rings in the disk, which we now know are quite common \citep[e.g.][]{2015ApJ...808L...3A, 2016ApJ...820L..40A, 2016PhRvL.117y1101I, 2018ApJ...863...44A, 2018ApJ...869L..41A}, may also impact the resonant encounters and therefore lend themselves as excellent topics for future study.

\section{Acknowledgements}
We wish to thank Chris Spalding, Alessandro Morbidelli, and Greg Laughlin for their helpful insights and comments. We also thank the referee for a useful report. S.M. is supported by the NSF Graduate Research Fellowship Program under Grant  DGE-1122492. K.B. is grateful to the David and Lucile Packard Foundation and the Alfred P. Sloan Foundation for their support.

\bibliographystyle{apj}
\bibliography{main_arXiv}

\end{document}